# Semileptonic $B \to D$ and $B \to D^*$ Decays from the Lattice [†]

**Laurent Lellouch** [‡] [§]

(UKQCD Collaboration)

CPT, CNRS Luminy, Case 907, F-13288 Marseille, France [¶]

## Abstract

We obtain form factors relevant for semileptonic $B \to D$ and $B \to D^*$ decays from a quenched lattice QCD calculation. Our results enable us to test Heavy-Quark Symmetry, determine the Isgur-Wise function as well as obtain the Cabibbo-Kobayashi-Maskawa (CKM) matrix element $V_{cb}$ from a measurement, by the CLEO collaboration, of the differential decay rate for semileptonic $B \to D^*$ decays. We find that the Isgur-Wise function has a slope of $-(0.9 {}^{+4}_{-4})$ at zero recoil and that $|V_{cb}|K(1) = 0.037 {}^{+1}_{-1} {}^{+4}_{-2} (0.99/(1+\beta_{A_1}(1)))$, where the first set of errors is due to experimental uncertainties and the second to our errors on $\rho^2$. Here $K(1)$ and $\beta_{A_1}(1)$ denote respectively power and radiative corrections at zero recoil.

*to appear in the Proceedings of the XXXth Rencontres de Moriond*
*"QCD and High Energy Hadronic Interactions"*
*Les Arcs, France, March 1995*

[†]This research was supported by the UK Science and Engineering Research Council under grants GR/G 32779 and GR/J 21347.
[‡]email: lellouch@cpt.univ-mrs.fr
[§]I thank my colleagues from the UKQCD Collaboration and G. Martinelli for fruitful discussions.
[¶]Unité Propre de Recherche 7061.

# 1 Introduction

Semileptonic $B \to D$ and $B \to D^*$ decays are interesting phenomenologically because their study enables one to determine the CKM matrix element $V_{cb}$ as well as to test QCD in its non-perturbative domain. At the level of quarks, the $b$ quark within the $B$ meson decays into a $c$ quark through the emission of a $W$ boson. The amplitude for these decays is proportional to $V_{cb}$ but the coupling of the quarks to the $W$ is significantly altered by the non-perturbative strong-interaction dynamics which binds the quarks into their respective mesons. It is to compute these corrections that we resort to lattice QCD.

The study of these decays is also interesting theoretically because it permits one to test the range of applicability of Heavy-Quark Symmetry (HQS). In a hadron composed of a heavy quark and light hadronic degrees of freedom one finds that the dynamics of the later become independent of the heavy quark's spin and mass when this mass is much larger than $\Lambda_{\rm QCD}$. Thus, to the extent that $m_b$, $m_c \gg \Lambda_{\rm QCD}$, an $SU(4)$ symmetry on the multiplet ($c \uparrow, c \downarrow, b \uparrow, b \downarrow$) emerges. This symmetry, of course, leads to simplifications but it also determines the dependence of physical quantities on heavy-quark mass. Now, because the mass and spin of a heavy quark can be varied almost at will in lattice calculations, they are ideal for testing HQS.

The QCD matrix elements required to describe these semileptonic decays can be parametrized in terms of six form factors:

$$\frac{\langle D(v')|\bar{c}\gamma^\mu b|B(v)\rangle}{\sqrt{m_B m_D}} = (v+v')^\mu h_+(\omega) + (v-v')^\mu h_-(\omega) ,$$
$$\frac{\langle D^*(v',\epsilon)|\bar{c}\gamma^\mu b|B(v)\rangle}{\sqrt{m_B m_{D^*}}} = i\epsilon^{\mu\nu\alpha\beta}\epsilon^*_\nu v'_\alpha v_\beta\, h_V(\omega) , \quad (1)$$
$$\frac{\langle D^*(v',\epsilon)|\bar{c}\gamma^\mu\gamma^5 b|B(v)\rangle}{\sqrt{m_B m_{D^*}}} = (\omega+1)\epsilon^{*\mu} h_{A_1}(\omega) - \epsilon^* \cdot v \left(v^\mu h_{A_2} + v'^\mu h_{A_3}\right) ,$$

where $\omega = v\cdot v'$ and $\epsilon^\mu$ is the polarization vector of the $D^*$. In the heavy-quark limit, these six form factors reduce to a single universal function, $\xi(\omega)$, known as the Isgur-Wise function and normalized to 1 at $\omega=1$ [2]. We have

$$h_i(\omega) = (\alpha_i + \beta_i(\omega) + \gamma_i(\omega))\, \xi(\omega) , \quad (2)$$

where $\alpha_+ = \alpha_V = \alpha_{A_1} = \alpha_{A_3} = 1$ and $\alpha_- = \alpha_2 = 0$. The functions $\beta_i$ parametrize perturbative, radiative corrections to the symmetry limit which we calculate using the results of [3]. The functions $\gamma_i$ parametrize non-perturbative corrections which correspond to matrix elements of higher-dimension operators in Heavy-Quark Effective Theory and which are proportional to inverse powers of the heavy-quark masses. Luke's theorem [4] guarantees that $\gamma_{+,A_1}(1)$ begins at second order in $(1/2m_{b,c})$.

The results for the dominant form factors, $h_+$ and $h_{A_1}$, presented below were obtained from 60 quenched configurations on a $24^3 \times 48$ lattice at $\beta=6.2$, corresponding to an inverse lattice spacing of approximately 2.85 GeV. Quark propagators were generated from an $\mathcal{O}(a)$-improved Wilson action [5] for three values of the light-quark mass around that of the strange and four values of the initial and final heavy-quark masses around that of the charm; the $b$ cannot be simulated directly because its mass is of the same order as our cutoff. To obtain results relevant for the $b$, then, we extrapolate "charm"-quark results as described below. Also, where necessary, we extrapolate the results linearly in light-quark mass to the chiral limit. (For details, see Ref. [1].)

## 2 Tests of Heavy Quark Symmetry

To extrapolate $h_+$ and $h_{A_1}$ obtained in the region of the charm to a domain relevant for $B$-decays we must understand these form factors' dependence on heavy-quark mass. For this purpose, we study the quantities $\bar{h}_+(\omega)/(1+\beta_+(\omega))$ and $\bar{h}_{A_1}(\omega)/(1+\beta_{A_1}(\omega))$ where we have subtracted all mass dependence due to radiative corrections. We work with $\bar{h}_i(\omega)\equiv h_i(\omega)/h_i(1)$, $i=+,A_1$, instead of $h_i(\omega)$ to reduce cutoff effects. Plotted in Fig. 1 is $\bar{h}_+/(1+\beta_+)$ for four different values of the heavy-quark mass for transitions where initial and final heavy quarks are degenerate in mass.

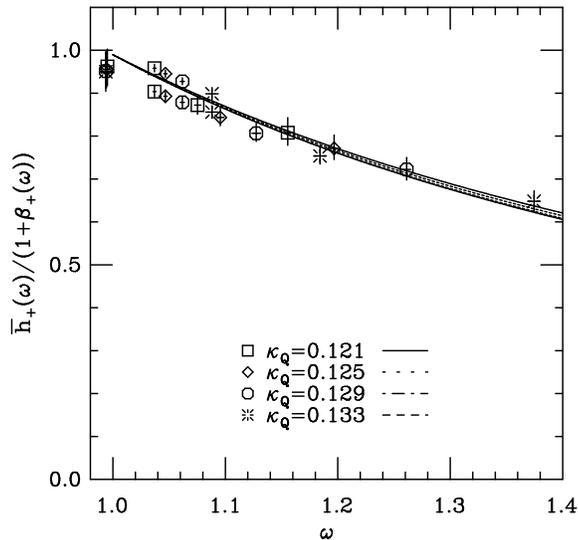
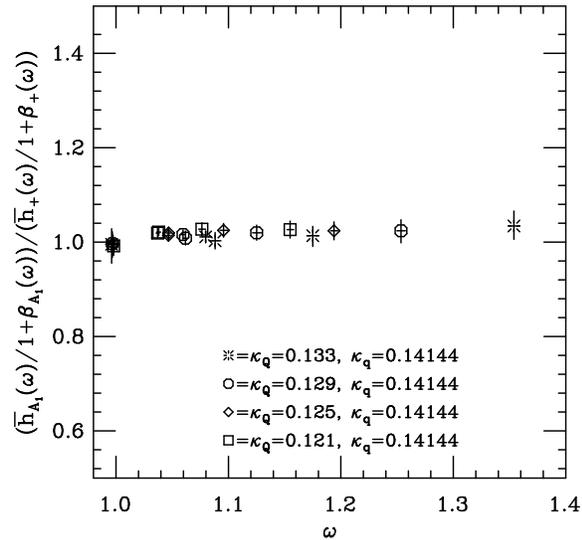

Figure 1: $\bar{h}_+/(1+\beta_+)$ for four sets of heavy-quark mass. Each set is individually fitted to the parametrization of Eq. (3) (curves).

Figure 2: Ratio $(\bar{h}_+/(1+\beta_+)/\bar{h}_{A_1}/(1+\beta_{A_1}))$. In both these figures, the light antiquark has a mass around that of the strange.

The fact that all four sets of points lie very much on the same curve indicates that power corrections to $\bar{h}_+$ are small over a large range of recoils, up to small residual cutoff effects. Thus, to a very good approximation, $\bar{h}_+/(1+\beta_+)$ is an Isgur-Wise function.

Nearly identical results obtain for $\bar{h}_{A_1}/(1+\beta_{A_1})$ and we are entitled to conclude that this quantity is the same Isgur-Wise function as the one given by $\bar{h}_+/(1+\beta_+)$. This is seen explicitely in Fig. 2 where we plot the ratio of $(\bar{h}_{A_1}/(1+\beta_{A_1}))/(\bar{h}_+/(1+\beta_+))$ for the same degenerate transitions as in Fig. 1.

The near equality of $\bar{h}_+/(1+\beta_+)$ and $\bar{h}_{A_1}/(1+\beta_{A_1})$ and their independence on heavy-quark mass is in stark contrast with the symmetry violations found in leptonic decays of $D$ mesons which are on the order of 30 to 40% [6]. The protection afforded by Luke's theorem at $w=1$ appears to extend over the whole range of recoils.

## 3 The Isgur-Wise Function and $|V_{cb}|$

Having established that the quantities $\bar{h}_+/(1+\beta_+)$ and $\bar{h}_{A_1}/(1+\beta_{A_1})$ are, to a good approximation, independent of heavy-quark mass, we can combine the results we have for different initial and final heavy-quark-mass and extrapolate them in light-quark mass to the chiral limit, obtaining in this way the Isgur-Wise function, $\xi(\omega)$, relevant for semi-leptonic $B \to D$ and $B \to D^*$ decays. Because our results for $\bar{h}_+/(1+\beta_+)$ are more accurate, we use that quantity alone to determine $\xi(\omega)$. We fit $\xi(\omega)$ to $s\xi_{NR}(\omega)$ where $s$ is introduced to absorb uncertainties in overall normalization and $\xi_{NR}(\omega)$ is a standard form for the Isgur-Wise function

$$\xi_{NR}(\omega) = \frac{2}{\omega+1}\exp\left[-(2\rho^2-1)\frac{\omega-1}{\omega+1}\right] \qquad (3)$$

which parametrizes $\xi$ in terms of the slope parameter $\rho^2 = -\xi'(1)$. We find

$$\rho^2 = 0.9\,{}^{+\,2}_{-\,3}\,{}^{+\,4}_{-\,2}, \qquad (4)$$

and $s = 0.96\,{}^{+\,2}_{-\,2}\,{}^{+\,5}_{-\,3}$, where the first errors are statistical and the second, systematic. This result for $\rho^2$ is in good agreement with earlier propagating-heavy-quark, lattice results [7].

In Fig. 3 we compare our Isgur-Wise function with $|V_{cb}|(1+\beta_{A_1}(1))K(\omega)\xi(\omega)$ measured by CLEO [8]. Agreement is excellent. Here $K(\omega)$ parametrizes radiative corrections away from $\omega=1$ as well as power corrections. If we assume that $K(\omega)$ is $\omega$-independent, which is reasonable given our errors on $\rho^2$, a fit of the CLEO data to $|V_{cb}|\xi_{NR}(\omega)$ with $\rho^2$ constrained to the lattice value of Eq. (4) yields

$$|V_{cb}|K(1) = 0.037\,{}^{+\,1}_{-\,1}\,{}^{+\,4}_{-\,2}\,(0.99/(1+\beta_{A_1}(1))), \qquad (5)$$

where the first set of errors is due to experimental uncertainties and the second, to our errors on $\rho^2$. To obtain $|V_{cb}|$, one must use estimates of the non-perturbative correction K(1). They range from 0.91(3) [9] to 0.945(25) [10].

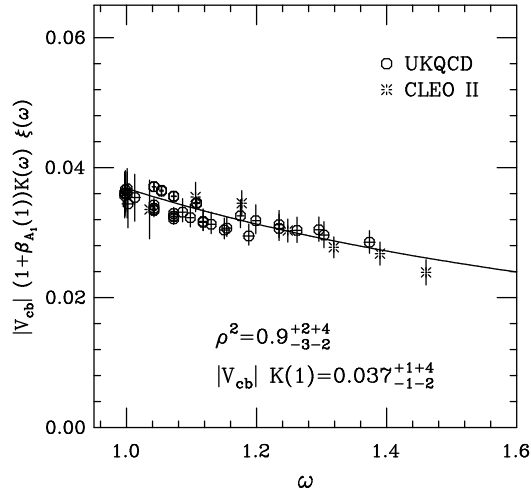

Figure 3: $|V_{cb}|(1+\beta_{A_1}(1)K(\omega)\xi(\omega)$ from CLEO (bursts) compared to our result for the Isgur-Wise function rescaled by $|V_{cb}|$ (circles) (see text).